\documentclass[apjl]{emulateapj}

\slugcomment{ApJ Letters, in press}

\shorttitle{Planet-planet scattering in planetesimal disks}
\shortauthors{Raymond, Armitage \& Gorelick}

\begin{document}

\title{Planet-planet scattering in planetesimal disks}

\author{Sean N. Raymond\altaffilmark{1,2}, Philip
  J. Armitage\altaffilmark{2,3} and  Noel Gorelick\altaffilmark{4}}
\altaffiltext{1}{Center for Astrophysics and Space Astronomy, 389 UCB,
  University of Colorado, Boulder CO 80309; sean.raymond@colorado.edu}
\altaffiltext{2}{Department of Astrophysical and Planetary Sciences, University of Colorado, Boulder CO 80309}
\altaffiltext{3}{JILA, 440 UCB, University of Colorado, Boulder CO 80309}
\altaffiltext{4}{Google, Inc., 1600 Amphitheatre Parkway, Mountain View, CA 94043}

\begin{abstract}
We study the final architecture of planetary systems that evolve under the
combined effects of planet-planet and planetesimal scattering.  Using
N-body simulations we investigate the dynamics of marginally unstable
systems of gas and ice giants both in isolation and when the planets form
interior to a planetesimal belt. The unstable isolated systems evolve under
planet-planet scattering to yield an eccentricity distribution that matches
that observed for extrasolar planets. When planetesimals are included the
outcome depends upon the total mass of the planets. For $M_{\rm tot}
\gtrsim 1 \ M_J$ the final eccentricity distribution remains broad, whereas
for $M_{\rm tot} \lesssim 1 \ M_J$ a combination of divergent orbital
evolution and recircularization of scattered planets results in a
preponderance of nearly circular final orbits. We also study the fate of
marginally {\em stable} multiple planet systems in the presence of
planetesimal disks, and find that for high planet masses the majority of
such systems evolve into resonance. A significant fraction lead to resonant
chains that are planetary analogs of Jupiter's Galilean satellites. We
predict that a transition from eccentric to near-circular orbits will be
observed once extrasolar planet surveys detect sub-Jovian mass planets at
orbital radii of $a \simeq 5-10 \ {\rm AU}$.
\end{abstract}


\section{Introduction}
Different dynamical mechanisms are commonly invoked to explain the
architecture of the outer Solar System and extrasolar planetary systems. In
the Solar System, scattering of small bodies (``planetesimals") by the ice
giants \citep{fernandez84,ida00,kirsh09} is thought to drive outward
planetary migration and concomitant capture of Pluto and other Kuiper Belt
Objects into resonance \citep{malhotra95,murrayclay05}. The effects of
planetesimal scattering on the gas giants are smaller but still
significant, for example in the ``Nice model" \citep{tsiganis05,gomes05}
where a divergent resonance crossing between Jupiter and Saturn triggers
the Late Heavy Bombardment.  The presence of small bodies around other
stars can be inferred from observations of debris disks \citep{wyatt08},
but as yet there is no evidence for a dynamical role of planetesimals in
known extrasolar planetary systems. At radii where tidal effects are
negligible (roughly $a \gtrsim 0.1$~AU) the eccentricity distribution of
extrasolar planets matches relatively simple models of gravitational
scattering among a system of two or more massive planets that typically
include neither planetesimals nor residual gas
\citep{chatterjee08,juric08}.

The success of pure planet-planet scattering models does not imply that 
other dynamical processes can be ignored. The observed distribution of 
semi-major axes of extrasolar planets at small orbital radii requires 
the existence of an additional dissipative process \citep{adams03}, 
most probably gas disk migration \citep{lin86}, which will itself affect 
planetary eccentricity \citep{moorhead05}. At larger orbital radii 
simple arguments suggest that a dynamically significant external
reservoir of planetesimals ought to be a common feature of young planetary
systems. The formation of giant planets becomes increasingly difficult
at large radii \citep{pollack96,kokubo02}, and hence it is probable that
disks of leftover debris surround the zone of giant planet formation in
most young systems. The typical masses of planetesimal disks are unknown,
but values of 30-50~$M_\oplus$ that are comparable to those inferred for
the early outer Solar System are plausibly typical, since they are
consistent with disk masses estimated from astronomical observations of the
youngest stars \citep{andrews05}. The dynamical effect of such disks on 
{\em currently} observed extrasolar planetary systems would be small, since 
radial velocity surveys preferentially detect planets that are either 
massive (and hence largely immune to influence from planetesimal disks) 
or orbit at very small radii where the mass of leftover debris is negligible.

Pooling knowledge from the Solar System and extrasolar planetary systems
motivates consideration of a model in which planet formation typically
yields a marginally unstable system of massive planets in dynamical contact
with both a residual gas disk and an exterior planetesimal disk. In this
{\em Letter} we ignore the gas disk and study the subsequent evolution
under the combined action of planet-planet and planetesimal scattering. We
do not attempt to model the full distribution of extrasolar planetary
properties (which would require the inclusion of hydrodynamic effects), but
rather focus on how planetesimal disks affect the final eccentricity of
extrasolar planets at moderately large orbital radii.

\section{Methods}
We assume that the gas-dominated epoch of planet formation is sufficiently
distinct from the subsequent phase of planet-planet and planetesimal
scattering that it makes sense to study the latter with pure N-body
simulations. We focus on two large ensembles of runs. The {\tt highmass}
set comprises 1000 integrations of three planet systems in which the masses
of the planets are chosen randomly in the range $M_{\rm Sat} < M_p < 3
M_J$, with a distribution,
\begin{equation}
 \frac{{\rm d}N}{{\rm d}M} \propto M^{-1.1},
\end{equation} 
which matches that observed \citep{marcy08}. The observed distribution is
derived from an incomplete sample that represents (in the context of our
model) the distribution {\em after} scattering, but these subtleties do not
matter for our purposes. The {\tt lowmass} set is identical except that we
sample a wider swath of the mass function between $10 M_\oplus$ and $3
M_{\rm J}$.  The planets are initially placed in a marginally unstable
configuration defined by circular, nearly coplanar orbits with a separation
of 4-5~$r_{h,m}$, where the mutual Hill radius,
\begin{equation}
 r_{h,m} = \frac{1}{2} \left( \frac{M_1+M_2}{3M_\star} \right)^{1/3} \left( a_1 + a_2 \right).
\end{equation} 
Here $a_1$ and $a_2$ are the planets' semi-major axes, $M_1$ and $M_2$
their masses, and $M_\star$ is the stellar mass. With this spacing the
instability timescale is relatively long \citep{chambers96,chatterjee08}
(the median timescale before the first planet-planet encounter was 0.3~Myr
for the {\tt highmass} integrations without disks). Our initial conditions
are only a small subset of the architectures predicted from giant planet
formation models \citep{thommes08,mordasini09}, though broadly consistent
with scenarios in which giant planets are captured into mean-motion
resonances during the late stages of gas disk evolution \citep{thommes08b}
prior to being removed from resonance by turbulent perturbations
\citep{adams08}.  Each integration is repeated twice, once with just the
three planets\footnote{Note that the simulations without planetesimal disks
are identical to the {\tt mixed1} and {\tt mixed2} cases from
\citep{raymond08,raymond09}. } and once with an external planetesimal disk
whose inner radius of $a_{\rm in} = 10 \ {\rm AU}$ is 2 Hill radii beyond
the orbit of the outermost planet\footnote{In the Nice model, a separation
of $\approx 3-4$~Hill radii between Neptune and the outer planetesimal disk
is needed to match the timing of the Late Heavy Bombardment
\citep{gomes05}.  The spacing of 2 $r_{h,m}$ means that our models evolve
on a somewhat shorter time scale.}. The inner edge of the disk lies within
the radius where a test particle in the restricted 3-body problem would be
stable, so the disk is in immediate dynamical contact with the outer
planet. The planetesimal disk is represented by 1000 bodies distributed
between 10 and 20~AU with a $\Sigma_{\rm disk} \propto r^{-1}$ surface
density profile and a total mass of $50 M_\oplus$.

We integrate these systems using the {\sc MERCURY} code \citep{chambers99}
for 100~Myr. The integrator uses the symplectic Wisdom-Holman mapping
\citep{wisdom91} for well-separated bodies, and the Bulirsch-Stoer method
when objects are within $N$ mutual Hill radii, where $N=3$ for our
case. Planets were removed if their orbital distances were smaller than 0.1
AU (``hit Sun'') or exceeded 100 AU (``ejection'').  Collisions were
treated as inelastic mergers conserving linear momentum.

A large ensemble of simulations includes some cases that are much harder to
integrate accurately than the majority. To make the best use of our
computational resources we adopted a default timestep (20~days) that
results in accurate integrations (as measured by the fractional orbital
energy conservation ${\rm d}E/E$) for the typical case. We then identified
those runs (about 10\%) in which energy was not adequately conserved and
re-ran them with a smaller timestep.  For runs without disks we re-ran
cases with ${\rm d}E/E > 10^{-4}$ with a timestep of 5~days, while for the
runs with disks we re-computed cases with ${\rm d}E/E > 5 \times 10^{-4}$
with a timestep of 10~days.  A small number of the re-run simulations
(typically 15-35) still did not meet our energy criterion and were
discarded.

\section{Results for marginally unstable planetary systems}
In the absence of planetesimal disks our model planetary systems are
typically unstable on Myr time scales. There are also systems that are
stable over the 100~Myr duration of our runs. In our initial analysis we
{\em assume} that the typical outcome of giant planet formation is a system
that, in the absence of a disk, would be unstable. We therefore analyze the
subset of disk-less simulations that are unstable, and compare the results
to the matched set of simulations that include disks. This is not a perfect
one-to-one comparison, since the chaotic nature of the evolution means that
disk-less planetary systems can display different instability time scales
in the presence of even negligible perturbations. Nonetheless we {\em do}
observe statistical differences between the evolution of systems (with
disks) that correlate with the stability of the disk-free systems, and
hence it makes sense as a first approximation to separately consider the
results for stable and unstable cases.

\begin{figure}
\plotone{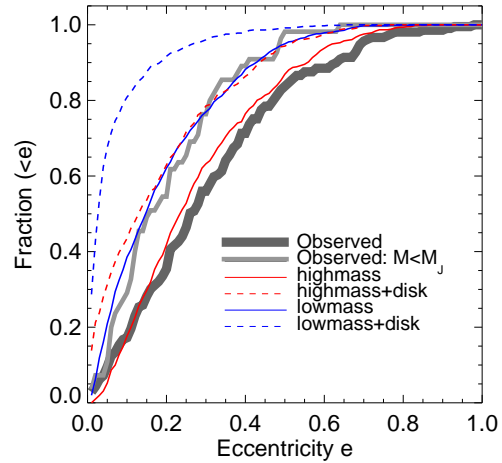}
\caption{Cumulative eccentricity distributions for observed extra-solar
  planets (thick grey line; lighter grey for minimum masses $M<M_J$) as
  compared with distributions from our simulations. The model distributions 
  include only the eccentricity of the innermost (and hence most easily 
  detected) planet.}
\label{fig:ecum}
\end{figure}

The results of our disk-less simulations agree with prior studies
\citep{chatterjee08,juric08,ford08}. Scattering from initially unstable
initial conditions frequently leads to the loss of one or more planets via
ejection or collisions and sets up a broad eccentricity distribution
\citep{rasio96,weidenschilling96,lin97}. Scattering among equal-mass
planets produces larger eccentricities than scattering of unequal-mass
planets \citep{ford03,raymond08}. Figure~\ref{fig:ecum} compares the final
eccentricities obtained from the unstable {\tt highmass} simulations and
the observed distribution \citep{butler06,schneider} of extra-solar
planets. They are in good quantitative agreement. The eccentricity
distribution from the unstable {\tt lowmass} simulations without disks is
shifted toward lower values \citep{raymond08,ford08}, and fits the observed
distribution of extra-solar planets with $M_p < M_J$ \citep{wright08}.  Our
model therefore exhibits evolution that is consistent with current
observations of extrasolar planetary systems, which as we noted previously
are mostly of systems at such small radii that planetesimal disks are
dynamically unimportant.

At larger radii we expect that both disks and planet-planet scattering will
play a dynamical role. A wide range of outcomes is then possible. Exchange
of energy and angular momentum between the planets and the planetesimal
disk leads to planetary migration
\citep{fernandez84,murray98,ida00,kirsh09}, which can be either stabilizing
or destabilizing. A low mass planet adjacent to the disk scatters
planetesimals inward, resulting in divergent migration that is often
stabilizing unless resonance crossing excites eccentricity to the point of
triggering instability.  Alternatively, an outer massive planet interacting
with the disk directly ejects planetesimals and migrates inward,
compressing the system and leading to instability or resonant capture. An
equally important effect is that the disk can act to recircularize the
orbits of scattered planets {\em after} dynamical instabilities
\citep{thommes99,ford07}.  To illustrate how significant recircularization
can be we ran a small additional set of idealized experiments in which a
single planet with mass $M_p$ on an orbit with $a = 10 \ {\rm AU}$ and $e =
0.5$ begins to interact with our {\em initial} planetesimal disk.  For
$10^4-10^5$ years (longer for smaller $M_p$), $e$ is damped roughly
exponentially with a damping time scale $t_e$, defined via,
\begin{equation}
 \frac{1}{e} \frac{{\rm d}e}{{\rm d}t} \equiv - \frac{1}{t_e}
\end{equation} 
of $t_e \approx 5 \times 10^4$ years, independent of $M_p$.  The subsequent
evolution was highly mass-dependent: for low-mass planets, $e$ continued to
decrease on much longer timescales (0.36, 0.63, and 4.6 Myr to reach $e
\lesssim 0.1$ for $M_p = 10 M_\oplus , 30 M_\oplus$, and $M_S$,
respectively).  Massive planets disrupted the disk, halting dynamical
friction.  The total decrease in $e$ for $M_p \gtrsim M_J$ was 0.15 or
less, corresponding to an increase of 1.5 AU or less in perihelion
distance.

\begin{figure*}
\plotone{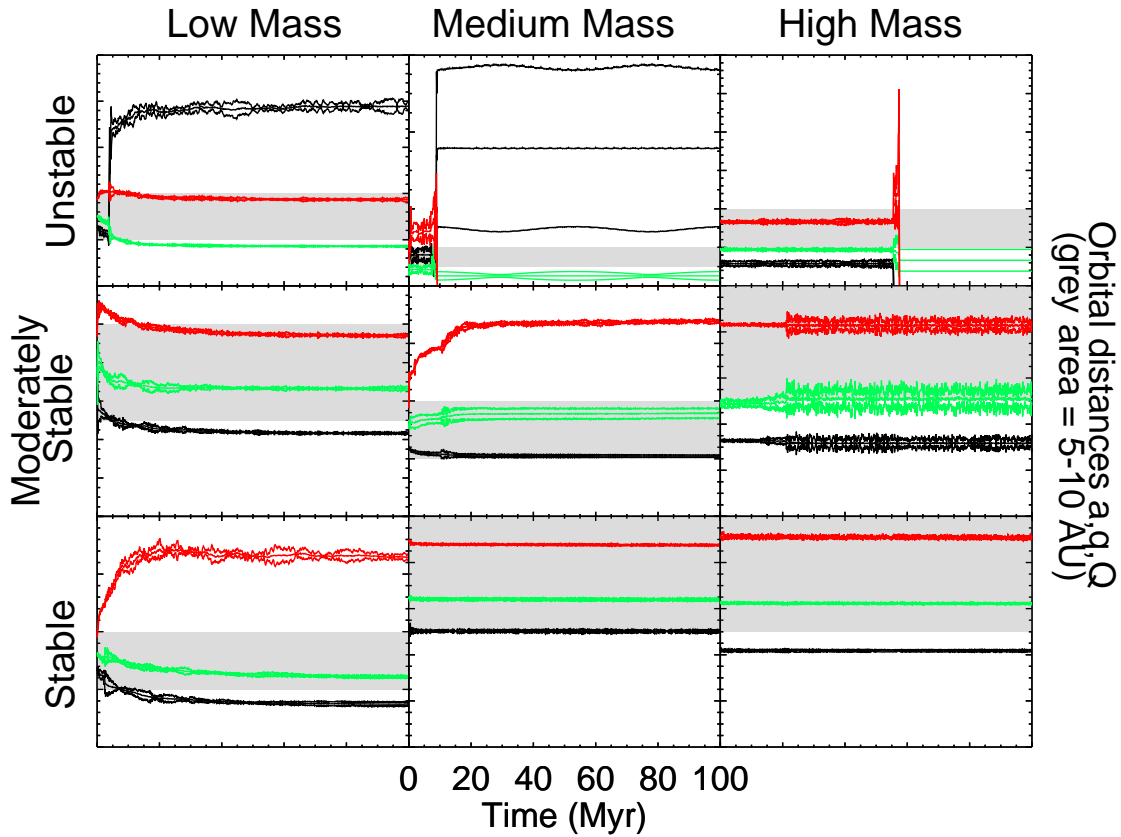}
\vspace{-0.3truein}
\caption{Evolution of a range of planetary systems interacting with
  planetesimal disks.  Each panel shows the evolution of the semimajor axis
  $a$, perihelion and aphelion distances $q$ and $Q$ for the three planets of
  a given simulation (planetesimal particles are not shown).  All simulations
  have the same $x$ axis scale, but different simulations have different $y$
  axis scales, so the region from 5-10 AU is shaded in grey for each case.
  Each column shows simulations in a given mass range: low-mass (total initial
  planet mass $M_{tot} < 0.5 M_J$), medium-mass, ($0.5 M_J < M_{tot} < 2
  M_J$), and high-mass ($M_{tot} > 2 M_J$).  Each row groups simulations by
  outcome: unstable cases underwent close encounters between planets, 
  moderately stable systems underwent significant orbital
  changes during the simulation but the system remained stable, and stable
  systems did not undergo any close encounters between planets and remained
  stable throughout.  The evolution in the center and top-left panels are
  qualitatively similar to different models of early Solar System
  dynamics \citep{thommes99,tsiganis05,gomes05}.}
\label{fig:evol9}
\end{figure*}

Figure~\ref{fig:evol9} illustrates the diversity of outcomes from our
simulations that include planetesimal disks. We split our simulations into
three mass bins (the Solar System's giant planets fall into the middle bin)
and three stability categories. In ``stable" systems there are no close
encounters between planets and no large-scale change in system architecture
(the ordering of the planets is preserved and all planets survive).
``Moderately stable" systems experience substantial perturbations -- which
may be due to resonance crossing in high-mass systems or close encounters
in low-mass systems -- that are nonetheless insufficient to alter the
architecture. ``Unstable" systems undergo close encounters leading to
architectural change.

Subsets of our runs show dynamics analogous to that studied for the Solar
System and for extra-solar planetary systems. At high masses planetesimal
disks stabilize about 30\% of cases but planet-planet scattering leading to
the loss of one or more planets is still common.  Quantitatively, the
median eccentricity is reduced (Figure~\ref{fig:ecum}) but many highly
eccentric systems remain.  As planet masses decreases the dynamical
importance of planetesimals grows. For low-mass systems, the masses of the
planets and the planetesimal disk are comparable and planetesimal
scattering inevitably leads to migration. Divergent crossing of mean-motion
resonances (one example of which is shown in the center panel of
Figure~\ref{fig:evol9}) can result in abrupt changes to planetary
semi-major axis and eccentricity that qualitatively resemble those seen in
the Nice model \citep{tsiganis05,gomes05}.  We also see behavior that
resembles an alternative Solar System model in which Uranus and Neptune
formed in the Jupiter-Saturn region and were scattered outward
\citep{thommes99} (top left panel).  At the lowest masses even highly
unstable systems rarely destroy any planets because recircularization of
scattered planets is efficient (top center panel), though re-ordering of
planets is common.  In summary, dynamics characteristic of the outer Solar
System is common among low- to medium-mass planetary systems.

\begin{figure}
\plotone{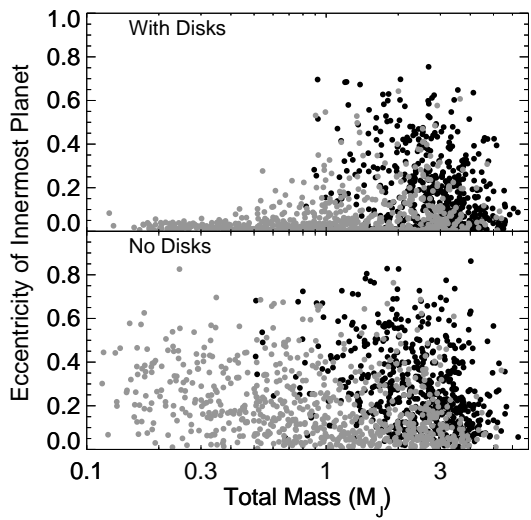}
\caption{Final eccentricity of the innermost planet as a function of the
  total mass in surviving planets for the {\tt highmass} (black) and {\tt
  lowmass} (grey) simulations. The plotted sample shows those systems that
  were unstable without disks (bottom panel), together with the matched
  sample including disks (upper panel). Disks result in a sharp transition
  to a low-eccentricity regime for system masses below about one Jupiter
  mass.}
\label{fig:mtot-e}
\end{figure}

The main prediction of our model is the statistical distribution of
planetary eccentricity as a function of planet mass.
Figure~\ref{fig:mtot-e} shows the distribution of the eccentricity of the
innermost surviving planet as a function of the total mass in surviving
planets.  In the absence of disks we observe similar behavior across all
system masses -- the shift to smaller eccentricities for the {\tt lowmass}
runs, seen in Figure~\ref{fig:ecum}, is not visually apparent. When disks
are included, the eccentricity distribution divides into two distinct
regimes: a low mass regime in which planetesimal dynamics dominates to
yield low eccentricities and a high mass planet scattering dominated regime
where planetesimals play a minor role.  For our specific parameters (inner
edge disk edge at 10~AU, and a disk mass of 50~$M_\oplus$ assumed to be
typical for a stellar metallicity $Z= Z_\odot$) the transition between
these regimes occurs for system masses $M_{\rm crit} \approx 1 \ M_J$.  We
predict that systems whose giant planets orbit between 5 and 10~AU, and
which have a total mass below 1~$M_J$, will typically have low eccentricity
orbits.  This critical mass should scale roughly linearly with the stellar
metallicity, as we expect the initial planetesimal disk mass to be
proportional to $Z$.  We expect the same qualitative behavior even if the
zone of giant planet formation extends to larger radii \citep{goldreich04},
though in this case the low eccentricity regime would only be observable
further out.

\section{Results for marginally stable planetary systems}
Although the eccentricity distribution of extrasolar planets is consistent
with the hypothesis that {\em all} newly formed multiple systems are
unstable in the absence of disks, this conclusion may also be biased by
selection effects.  Most known extrasolar planets probably suffered
significant gas disk migration prior to scattering
\citep{lin96,trilling98,bodenheimer00}, so the high incidence of
instability may be a consequence of migration rather than formation. With
this in mind we have separately analyzed those (previously excluded)
systems that were stable in the absence of disks to see what impact disks
have on them.  As expected, low eccentricity outcomes
predominate. Resonances are also common: about 70\% of all stable {\tt
highmass} simulations and 1/3 of stable {\tt lowmass} simulations include
at least one pair of planets in the 3:2 or, more often, the 2:1 mean motion
resonance. This is a much higher probability of resonance capture than
occurs for pure planet-planet scattering without disks \citep{raymond08},
and it also exceeds the fraction of resonant systems that are expected to
survive the gaseous disk phase in the presence of turbulence
\citep{adams08}. Most surprisingly within the {\tt highmass} set a
substantial fraction (about 1/3) of stable systems become locked into
mean-motion resonances that involve {\em all three} of the planets --
analogs of the Laplace resonance among Jupiter's Galilean satellites.
Chains of resonances\footnote{Our definition of resonance requires one
resonant argument to librate with an amplitude $A < 150^\circ$. Roughly
half of the resonant systems were deep in the resonance, with $A <
60^\circ$. This fraction appears to be independent of the number of planets
in resonance, as about 1/4 of the {\tt highmass} resonant chains had $A <
60^\circ$ for {\em both} pairs of planets.} arise preferentially in
higher-mass systems and in systems where planetesimal-driven migration
causes compression rather than divergent migration.  Detection of high mass
planets at the relevant radii (between 5-10~AU) should soon be possible via
astrometric or direct imaging techniques, and observation of resonant
chains would be consistent with our model.  Intriguingly, the
recently-discovered triple planet system HR 8799 may be in a 4:2:1 resonant
chain \citep{marois08,fab09}.  Determining whether capture into resonance
was initiated by a gas or planetesimal disk may be possible via detailed
comparison of the outcome of resonant capture in planetesimal
\citep{murrayclay05} versus gas disks \citep{lee02,adams08}.

\section{Conclusions}
Circumstantial evidence suggests that many observed properties of the outer
Solar System \citep{malhotra95} and of extrasolar planetary systems
\citep{rasio96,weidenschilling96,lin97} may be attributable to the
dynamical effects of planetesimal scattering and planet-planet
scattering. Here, we have studied the predicted architecture of planetary
systems that results from the joint action of both mechanisms.  We have
argued that this regime will be relevant once lower mass extrasolar planets
are discovered at larger orbital radii than those currently known.
Generically we predict that a transition to ``Solar-System-like"
architectures, characterized by near-circular orbits and relatively stable
planetary separations, will be observed once surveys detect planets in the
regime where planetesimal disks play a dynamical role.  Our simulations
suggest that the transition is a surprisingly sharp function of total
planetary system mass, and that it occurs for system masses a factor of
several larger than the initial planetesimal disk mass.

Our initial conditions do not sample anything approaching the full range of
initial planetary system architectures. We believe that the existence of a
transition between typically eccentric and near-circular orbits is a
general feature of joint models of planet-planet and planetesimal
scattering, but the transition mass and minimum orbital radii at which
planetesimal effects become manifest is of course a function of the poorly
known masses and radial extent of planetesimal disks. Our results suggest
that the transition might be seen for sub-Jovian mass planets at orbital
radii of 5-10~AU, but the transition would be pushed to greater orbital
radii if giant planet formation consumes planetesimals across a wider
extent of the disk. We also find that the final system architecture varies
substantially depending on the initial separation of the planets. In
particular, if planet formation yields a mixture of massive systems in
initially stable orbits, interaction with planetesimals drives a large
fraction of systems into resonance. Whether such systems exist should be
testable in the near future.

\acknowledgements

We thank Google for the large amount of computer time needed for these
simulations. S.N.R. acknowledges support from NASA's Astrobiology Institute
through the Virtual Planetary Laboratory lead team, and from NASA's Origins
of Solar Systems program (NNX09AB84G). P.J.A.  acknowledges support from
the NSF (AST-0807471), from NASA's Origins of Solar Systems program
(NNX09AB90G), and from NASA's Astrophysics Theory program (NNX07AH08G).

\end{document}